\documentclass[prb,twocolumn,amsmath,amssymb,citeautoscript,showkeys,floatfix,superscriptaddress]{revtex4-2}

\usepackage{chemformula} 
\usepackage[T1]{fontenc} 
\usepackage{siunitx}
\usepackage{amssymb}
\usepackage{amsmath}
\usepackage{graphicx}
\usepackage{multirow}
\usepackage{booktabs}
\usepackage{footnote}
\usepackage[flushleft]{threeparttable}
\usepackage{comment}

\usepackage{hyperref}
\hypersetup{
    colorlinks,%
    citecolor=blue,%
    linkcolor=blue,%
    urlcolor=blue
}

\newcommand{\orcid}[1]{\href{https://orcid.org/#1}{\includegraphics[width=8pt]{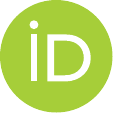}}}

\usepackage{filecontents}

\begin{document}


\title{Role of chalcogen vacancies and hydrogen in the optical and electrical properties of bulk transition-metal dichalcogenides}

\author{Shoaib Khalid\orcid{0000-0003-3806-3827}} 
\affiliation{Princeton Plasma Physics Laboratory, P.O. Box 451, Princeton, New Jersey 08543, USA}

\author{Anderson Janotti\orcid{0000-0002-0358-2101}}
$\email{janotti@udel.edu}$
\affiliation{Department of Materials Science and Engineering, University of Delaware, Newark, DE 19716, USA}

\author{Bharat Medasani\orcid{0000-0002-2073-4162}}
\affiliation{Princeton Plasma Physics Laboratory, P.O. Box 451, Princeton, New Jersey 08543, USA}


\begin{abstract}

Like in any other semiconductor, point defects in transition-metal dichalcogenides (TMDs) are expected to strongly impact their electronic and optical properties. However, identifying defects in these layered two-dimensional materials has been quite challenging with controversial conclusions despite the extensive literature in the past decade. Using first-principles calculations, we revisit the role of chalcogen vacancies and hydrogen impurity in bulk TMDs, reporting formation energies and thermodynamic and optical transition levels.  We show that the S vacancy can explain recently observed cathodoluminescence spectra of MoS$_2$ flakes and predict similar optical levels in the other TMDs. In the case of the H impurity, we find it more stable sitting on an interstitial site in the Mo plane, acting as a shallow donor, and possibly explaining the often observed n-type conductivity in some TMDs.  We also predict the frequencies of the local vibration modes for the H impurity, aiding its identification through Raman or infrared spectroscopy.

\end{abstract}


\maketitle

There is great interest in the development of transition-metal dichalcogenide (TMD) semiconductors for electronics~\cite{radisavljevic2011single,lopez2013ultrasensitive,late2013sensing, georgiou2013vertical,bernardi2013extraordinary,choi2012high,koperski2015single,kern2016nanoscale}. These materials have band gaps in the near infrared range, varying with the combination of metal(M) and chalcogen(X) atoms. Due to their layered structure, they can potentially be combined with other 2D materials such as hexagonal boron nitride or graphene forming van der Waals heterostructures, creating novel materials with properties suitable for applications in flexible electronics~\cite{lee2013flexible,geim2013van,li2016heterostructures,velicky2017two}. 
As semiconductors, understanding the behavior of native defects that can incorporated during growth or processing is 
an important step toward controlling the electrical conductivity in TMDs. Defects have been shown to affect carrier concentrations, scattering, and carrier mobilities~\cite{najmaei2015synthesis,zou2015open,nan2014strong}. More specifically, defect engineering has become prominent in altering the electrical~\cite{radisavljevic2011single}, optical~\cite{fai2010atom,di2012coupled,mak2013tightly}, and mechanical~\cite{bertolazzi2011stretching,castellanos2012elastic} behavior TMDs.

Identification of defects is the most important step in defect engineering. There are various techniques used to detect the defects including microscopy techniques such as scanning tunneling microscopy (STM),~\cite{liu2015line,barja2019identifying} atomic force microscopy (AFM),~\cite{barja2019identifying,rosenberger2018electrical} and optical spectroscopy techniques such as cathodoluminescence (CL), and photoluminescence (PL)~\cite{fabbri2016novel,yan2019tunable,verhagen2020towards,wu2017defect}. There is an added advantage in using optical techniques since they do not destroy the sample. Optical techniques involve creating electron-hole pairs and measuring the broad emission peak in the optical spectrum created by electrons in the defect states recombining with holes. First-principles calculations have been used previously to identify the defects by computing the optical emission from the defects and the sub-band levels created by the defects in the band gap for different materials~\cite{lyons2010carbon,frodason2020self,lyons2017first}. In this work, using first principle calculations we discuss the possible mechanism behind the near-infrared broad emission peak related to bulk MoS$_2$ recently measured by Fabri \textit{et al.}~\cite{fabbri2016novel}. We also calculate the emission energy corresponding to chalcogen vacancy for other members of the bulk TMD family. Combining the first principle calculations with optical spectroscopy techniques provides a powerful tool to identify individual point defects with great accuracy. 

\begin{figure*}
\includegraphics[width=12 cm]{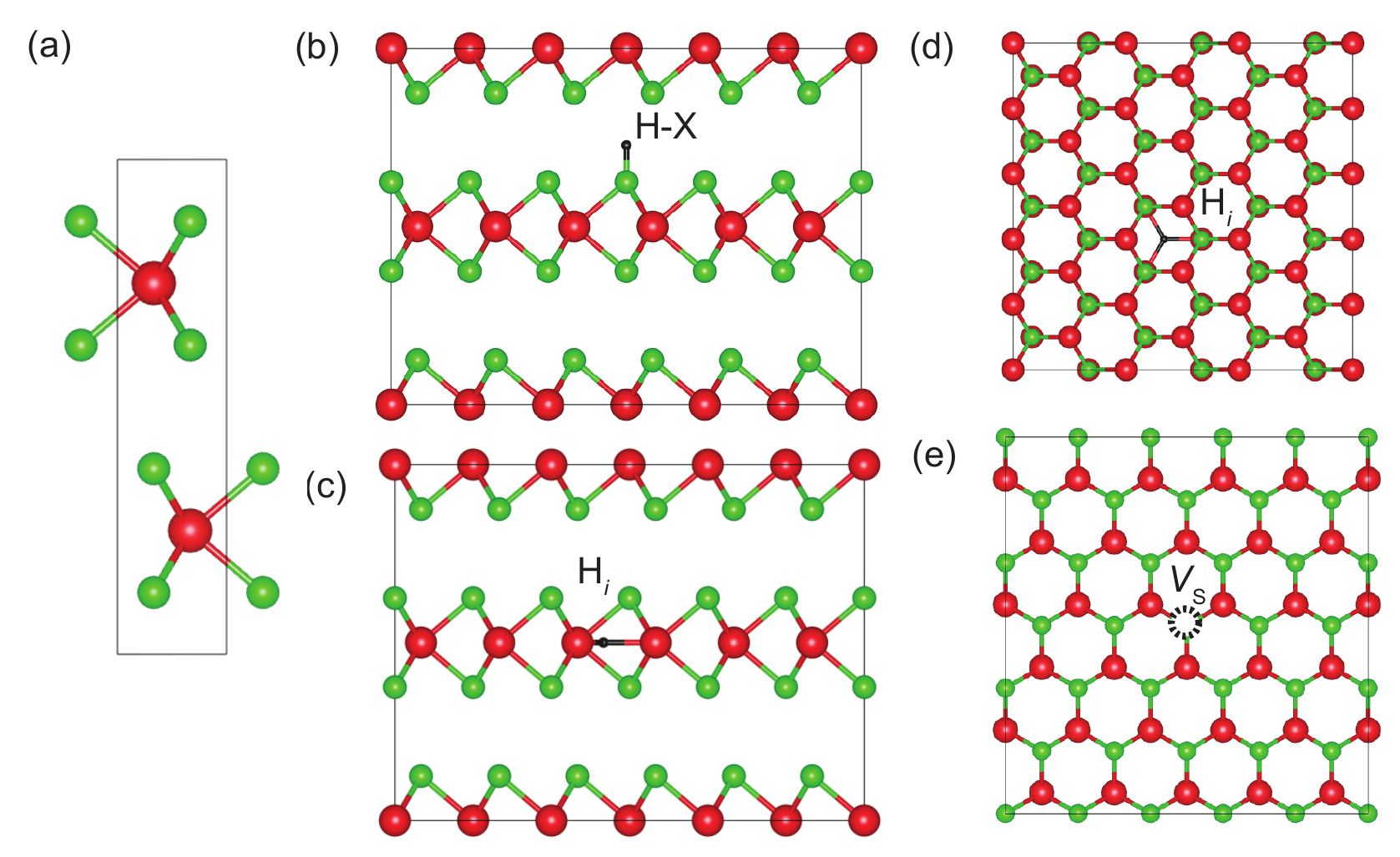}
\centering
\caption{(a) Primitive cell of bulk TMD's. The 180 atoms supercell used for defect calculations with (b) H-X defect(side view) (c) H$_i$ defect (side view) (d) H$_i$ defect (top view) and (e) chalcogen vacancy in the middle layer of the defect supercell (view from just above the middle layer).}
\label{fig1}
\end{figure*}

TMDs have also attracted attention as candidate materials for hydrogen storage applications~\cite{1972slsl.book,nikishenko1982quantum} because of their weak interlayer interaction and large interlayer spacing~\cite{makara1997hydrogen}. MoS$_2$ nanotubes have already shown promising signs as a hydrogen storage material~\cite{chen2001electrochemical,chen2003novel}. Hydrogen is also a commonly occurring impurity during the growth of TMDs, so understanding the possible configurations of hydrogen and its electronic behavior is important. To fully understand the behavior of H in TMDs, we studied H in several different configurations. We show that hydrogen interstitial, H$_i$, is the most stable configuration in bulk TMDs except for MoS$_2$. Vibrational frequencies related to H$_i$ in bulk TMDs are discussed and can provide direct evidence of the presence of H in these materials through Raman and IR spectroscopy measurements.

Our density functional calculations~\cite{hohenberg1964inhomogeneous,kohn1965self} are based on meta-GGA strongly constrained and appropriately normed (SCAN) functional~\cite{Jianwei2015}. The interaction between valence electrons and ionic cores is treated by projector augmented wave (PAW) potentials~\cite{blochl1994projector} as implemented in VASP code~\cite{kresse1993ab,kresse1994ab}. We used a plane wave cutoff of 400 eV with an energy convergence of 1 meV/atom. The Brillouin zone is sampled with 6$\times$6$\times$4 $\Gamma$-centered mesh. The van der Waals interaction between the layers was treated by empirical correction proposed by Grimme~\cite{grimme2010consistent}. For calculating the Huang-Rhys (HR) factors and total mass-weighted distortions, $\Delta$Q, we used the Nonrad code~\cite{turiansky2021nonrad}. For the defect calculations, orthogonal lattice vectors are used to make an 180 atoms supercell, as shown in  Fig.~\ref{fig1}. Spin polarization is included in cases with unpaired electrons. For the charge defect calculations, we used the recently developed self-consistent charge state correction (SCPC)~\cite{silva2021} as implemented in the VASP code.
The effect of spin-orbit coupling is also included.

\begin{table}
\begin{center}
\caption{{Lattice parameter and band gap of bulk TMDs calculated using SCAN. The experimental results are taken from \cite{kam1982detailed,lezama2014surface,murray1979thermal,fortin1982photovoltaic,Tan2020,james1963crystal}.}}
\begin{threeparttable}
\setlength{\tabcolsep}{6pt} 
\renewcommand{\arraystretch}{1.5} 
\begin{tabular}{lcccc}
  \toprule\toprule
  \multirow{2}{*}{\raisebox{-\heavyrulewidth}{Material}} & \multicolumn{2}{c}{Lattice parameter ({\AA}) } & \multicolumn{2}{c}{Band gap (eV)}\\
  \cmidrule(lr){2-5}
  & a=b, c & Exp. & E\textsubscript{g} & Exp. \\
  \midrule
  MoS\textsubscript{2} & 3.17, 12.45 & 3.15, 12.29 & 1.07 & 1.23,1.2,1.17  \\
  MoSe\textsubscript{2} & 3.29, 13.12 & 3.29, 12.93 & 1.03 & 1.09  \\
  MoTe\textsubscript{2} & 3.50, 14.26 & 3.51, 13.97 & 0.84 & 0.85  \\
  WS\textsubscript{2} & 3.16, 12.51 & 3.17, 12.36 & 1.17 & 1.35  \\
  WSe\textsubscript{2} & 3.28, 13.19 & 3.28, 12.95 & 1.13 & 1.20  \\
   \bottomrule\bottomrule
\end{tabular}
\label{table:formation2}
\end{threeparttable}
\end{center}
\end{table}

For bulk TMDs including  MoS$_2$, 2$H$ is the most stable phase. It is composed of $AB$ stacked atomic layers, where each layer forms a hexagonal lattice in the primitive cell as shown
in Fig.~\ref{fig1}. Each layer interacts with the neighboring layer by weak van der Waals forces. The equilibrium lattice parameters are obtained with six atoms primitive cell using SCAN functional with Grimme Van der Waals dispersion correction~\cite{grimme2010consistent}. In Table~\ref{table:formation2}, we list the equilibrium lattice parameters and band gap, which are in good agreement with the experimental data~\cite{james1963crystal,murray1979thermal}. In general, the in-plane and out-of-plane lattice parameters depend strongly on the chalcogen atom and very weakly on the metal atom. The band gap decreases as we move from the lighter S atom to the heavier Te atom.

\begin{figure*}
\includegraphics[width=15 cm]{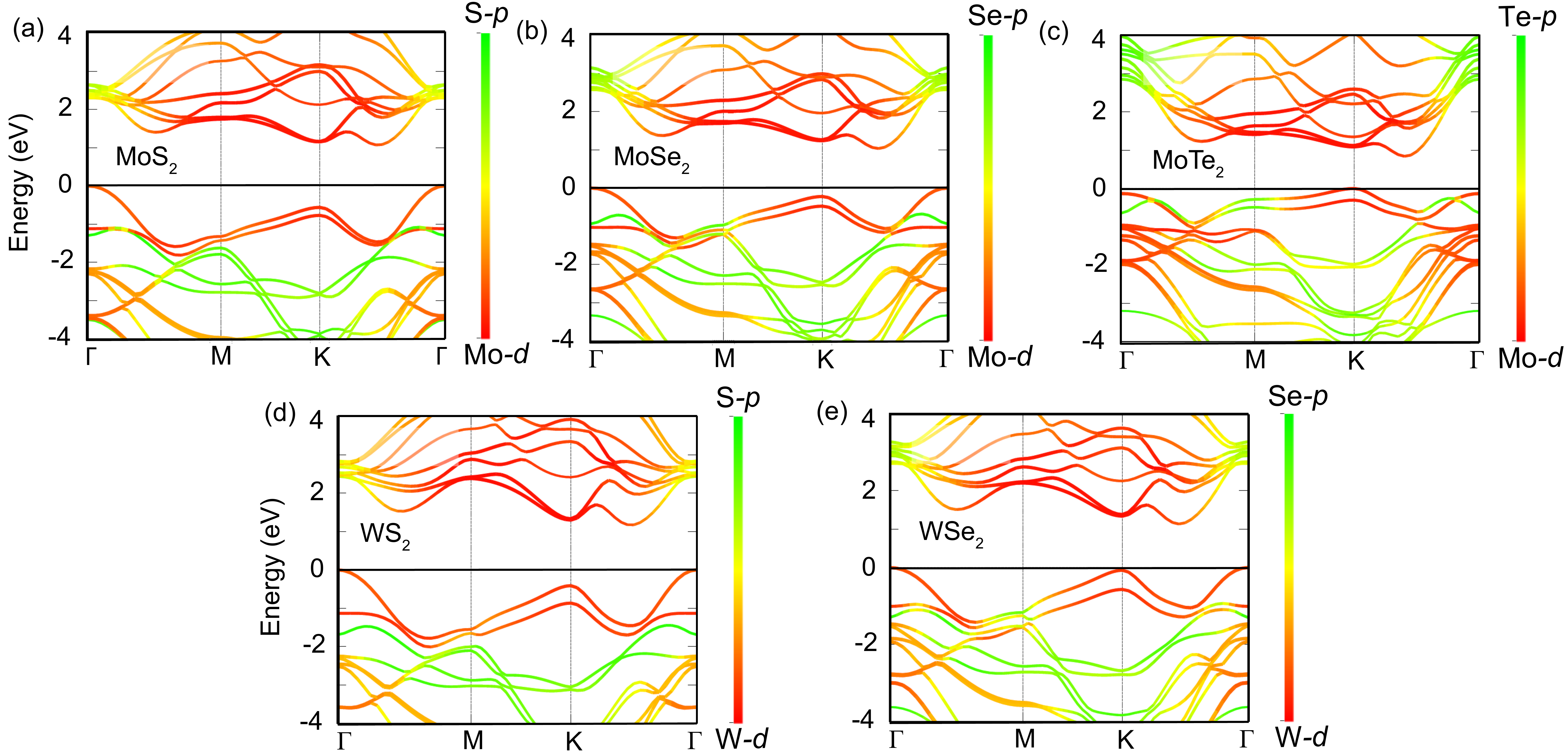}
\centering
\caption{Orbital-resolved electronic band structures of (a) MoS\textsubscript{2}, (b) MoSe\textsubscript{2}, (c) MoTe\textsubscript{2}, (d) WS\textsubscript{2} and (e) WSe\textsubscript{2} using the SCAN functional and including spin–orbit coupling. The valence band maximum (VBM) is set to zero in each case. Note that the band gap is indirect for all the studied bulk TMDs with VBM at $\Gamma$ and CBM lies in direction between K and $\Gamma$, for MoTe\textsubscript{2} the VBM is at K.}
\label{fig2}
\end{figure*}

In Fig.~\ref{fig2}, we show the orbital-resolved electronic band structure of bulk TMDs. All the studied TMDs are semiconductors with indirect band gaps; the valence-band maxima (VBM) occur at $\Gamma$ (except for MoTe$_2$ which has VBM at K) and the conduction-band minima (CBM) are along the high symmetry $\Gamma$-K direction. The spin-orbit splitting increases from the lighter S atom to the heavier Te atom for the same the metal atom. It is interesting to note that the band dispersion decreases from S to Te, such that the VBM shifts from $\Gamma$ to K in MoTe$_2$, whereas the position of CBM remains unchanged. The VBM and CBM have most of the contribution from metal $d$ states with some contribution from the chalcogen $p$ states. The calculated band gap using SCAN is in good agreement with the previous reports~\cite{Ivanovskaya2008,Padilha2014,komsa2015native,Tan2020} and experimental data~\cite{kam1982detailed,lezama2014surface,murray1979thermal,fortin1982photovoltaic}.

\begin{figure*}
\includegraphics[width=11 cm]{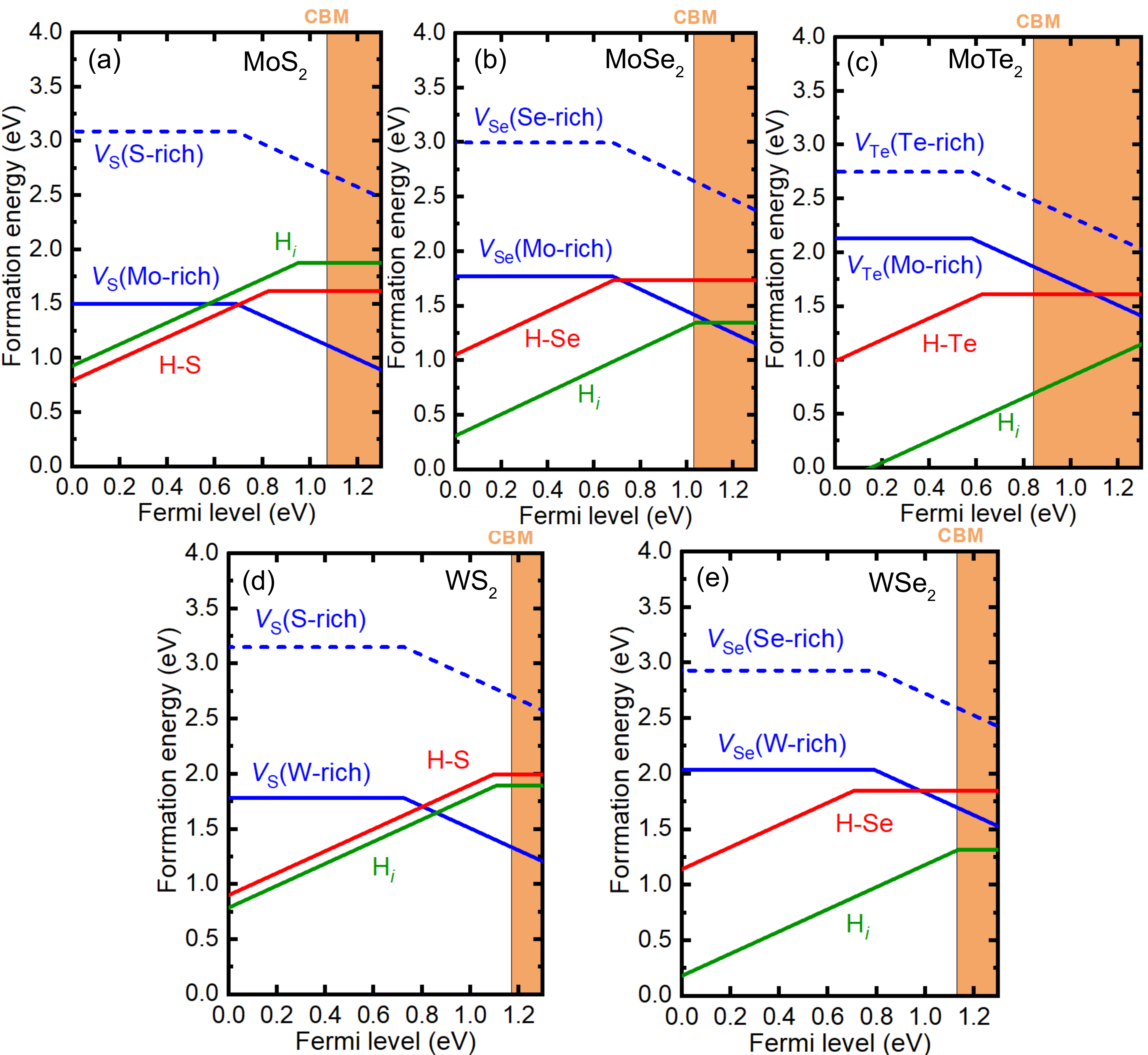}
\centering
\caption{Formation energies of chalcogen vacancy(dashed blues lines for chalcogen rich and solid blue lines for metal rich), H\textsubscript{$i$}(green lines) and H-X (X=chalcogen atom)(red lines)   as a function of Fermi level for (a) MoS\textsubscript{2}, (b) MoSe\textsubscript{2}, (c) MoTe\textsubscript{2}, (d) WS\textsubscript{2} and (e) WSe\textsubscript{2}. The VBM is used as a reference and the position of CBM is also shown in each case. }
\label{fig3}
\end{figure*}

The chalcogen vacancies have been observed by direct techniques such as high-resolution transmission electron microscopy~\cite{zhou2013intrinsic,lehtinen2015atomic,zhu2017defects,komsa2012,jeong2017heterogeneous,leiter2020situ} and also by indirect methods based on chemical reactions and x-ray photoelectron spectroscopy~\cite{qiu2013hopping,han2016photoluminescence,chen2015environmental,jeong2017heterogeneous,tosun2016air}. Komsa \textit{et al.}~\cite{komsa2015native} studied different native defects including vacancy complexes in MoS$_2$ and showed that the sulfur vacancy has the lowest formation energy.  In this work, we first focus on the chalcogen vacancy. The formation energy of the chalcogen vacancy as a function of Fermi level is shown in Fig.~\ref{fig3} for all the studied bulk TMDs. The chalcogen vacancy in bulk TMDs is a deep acceptor with the (0/-1) transition level deep inside the band gap in agreement with previous studies~\cite{komsa2015native,singh2022,kim2022experimental}. The relaxation around the vacancy is very minimal. The formation energy of neutral chalcogen vacancy in the chalcogen poor limit and the (0/-1) transition levels are shown in Table~\ref{table:transition2}. In the neutral charge state, the chalcogen vacancy leads to an occupied state just below the VBM and doubly degenerate empty states within the band gap. The degenerate states in the band gap are highly localized on neighboring metal atoms and can be occupied by four electrons. In the -1 charge state, one of the states is half occupied. The -2 charge state is unstable within the band gap for all the studied TMDs, in agreement with previous studies~\cite{komsa2015native,singh2022,kim2022experimental}.

\begin{table}
\begin{center}
\caption{{Formation energy of chalcogen vacancy in neutral charge state and (0/-1) transition level with respect to VBM for all the studied TMDs.}}
\begin{threeparttable}
\setlength{\tabcolsep}{6pt} 
\renewcommand{\arraystretch}{1.5} 
\begin{tabular}{lcc}
  \toprule\toprule
  \multirow{2}{*}{\raisebox{-\heavyrulewidth}{Material}} & \multicolumn{1}{c}{Formation energy ({eV}) } & \multicolumn{1}{c}{Transition level (eV)}\\
  \cmidrule(lr){2-3}
  & neutral &  (0/-1) \\
  \midrule
  MoS\textsubscript{2} & 1.50 & 0.68  \\
  MoSe\textsubscript{2} & 1.77& 0.68  \\
  MoTe\textsubscript{2} & 2.13 & 0.58  \\
  WS\textsubscript{2} & 1.78 & 0.73  \\
  WSe\textsubscript{2} & 2.03 & 0.79  \\
   \bottomrule\bottomrule
\end{tabular}
\label{table:transition2}
\end{threeparttable}
\end{center}
\end{table}

\begin{table}
\begin{center}
\caption{Calculated Huang-Rhys (HR) factors for initial state $S$\textsubscript{$i$} and final state $S$\textsubscript{$f$}, total mass-weighted distortions $\Delta$Q  and emission energy due to chalcogen vacancy for bulk TMD's.}
\begin{threeparttable}
\setlength{\tabcolsep}{3.8pt} 
\renewcommand{\arraystretch}{1.5} 
\begin{tabular}{lccccc}
  \toprule\toprule
  \multirow{2}{*}{\raisebox{-\heavyrulewidth}{Material}} 
  &\multicolumn{2}{c} {HR factor}  &\multicolumn{1}{c} {$\Delta$Q} & \multicolumn{1}{c}{$E$\textsubscript{PL}} & \multicolumn{1}{c}{$E$\textsubscript{FC}} \\
  \cmidrule(lr){2-3}
  & $S$\textsubscript{$i$} & $S$\textsubscript{$f$} & ($\sqrt{amu}$ {\AA}) & (eV) & (meV)\\
  \midrule
  MoS\textsubscript{2} & 2.73 & 1.42 & 1.05 & 0.62 & 68 \\
  MoSe\textsubscript{2} & 4.33 & 1.94 & 1.44 & 0.72 & 98  \\
  MoTe\textsubscript{2} & 6.14 & 3.33 & 2.07 & 0.61 & 103 \\
  WS\textsubscript{2} & 5.30 & 5.23 & 1.62 & 0.81 & 90\\
  WSe\textsubscript{2} & 7.54 & 8.34 & 2.17 & 0.97 & 108 \\
   \bottomrule\bottomrule
\end{tabular}
\label{table:lattice_parameter}
\end{threeparttable}
\end{center}
\end{table}

\begin{figure}
\includegraphics[width=8 cm]{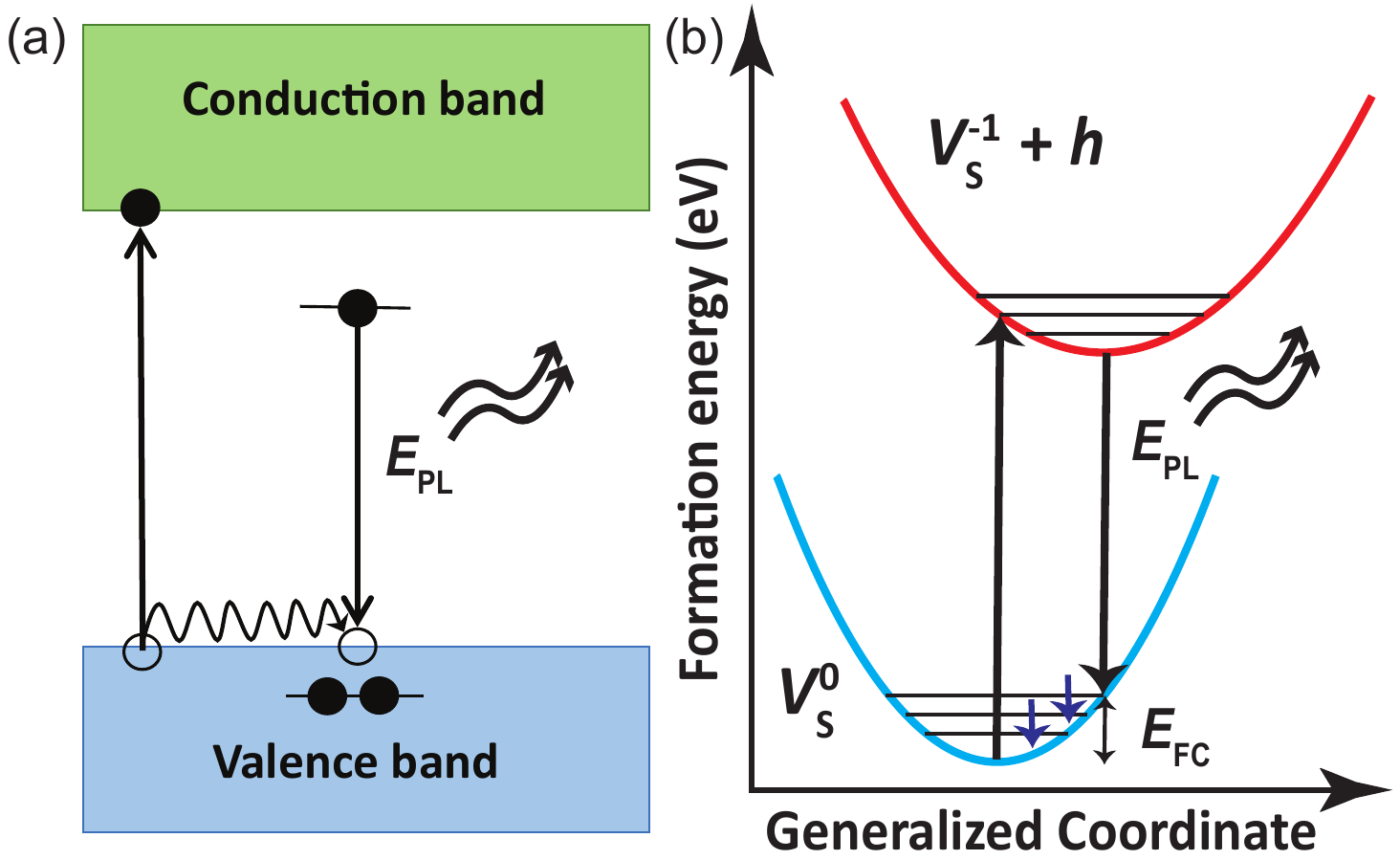}
\caption{(a) Schematic of optical emission due to chalcogen vacancy in bulk TMDs (b) Configuration coordinate diagram for the chalcogen vacancy in TMDs. The chalcogen vacancy in the neutral charge state is considered as the ground state, and the energy variation as a function of atomic displacement around the stable configuration is shown. The photoluminescence emission energy related to chalcogen vacancy is labeled as $E$\textsubscript{PL}.}
\label{fig4}
\end{figure}

Fabri \textit{et al.}~\cite{fabbri2016novel} in their cathodoluminescence experiment observed a near-infrared emission peak around 0.75 eV for exfoliated bulk MoS$_2$ flakes. Gupta \textit{et al.}~\cite{gupta2018franck} studied the Franck-Condon shift associated with monolayer MoS$_2$ and related the emission energy to the S vacancy in monolayer MoS$_2$ with the one measured by Fabri \textit{et al.}. It is important to mention here that the exfoliated MoS$_2$ flakes measured by Fabri \textit{et al.}~\cite{fabbri2016novel} lie in the bulk regime, and that exciton binding energies in the bulk are much smaller than the exciton binding energy in the monolayer MoS$_2$.  In this work, we studied the optical emission due to chalcogen vacancy in bulk MoS$_2$ and other members of the bulk TMD family. The schematic of optical emission due to chalcogen vacancy in bulk TMDs is shown in Fig.~\ref{fig4}(a). An electron-hole pair is generated in a photoluminescence (PL)/cathodoluminescence (CL) excitation. The generated electron in the conduction band does not recombine with the hole in the valence band, instead, the electron in the -1 defect state of chalcogen vacancy recombines with a hole in the valence band emitting a photon with energy $E$\textsubscript{PL}. In Table~\ref{table:lattice_parameter}, we list the emission energy, $E$\textsubscript{PL}, for all the studied TMDs. Our calculated emission energy for MoS$_2$ bulk is around 0.62 eV, in reasonable agreement with experimental observation \cite{fabbri2016novel}. In Fig.~\ref{fig4}(b), we show the schematic of the configuration coordinate diagram for the sulfur vacancy in neutral and -1 charge states; the electron in the -1 charge state combines with the hole in the neutral charge state emitting a photon with energy $E$\textsubscript{PL}. The Franck-Condon shift, $E$\textsubscript{FC}, which is the energy difference of $V$\textsubscript{S}\textsuperscript{0} in ground state configuration and  $V$\textsubscript{S}\textsuperscript{0} in $V$\textsubscript{S}\textsuperscript{-1} configuration, the total mass-weighted distortions, $\Delta$Q, and Huang-Rhys (HR) factors for initial state $S$\textsubscript{$i$} and final state $S$\textsubscript{$f$} are calculated and listed in Table~\ref{table:lattice_parameter}. $\Delta$Q corresponds to the difference in geometry between the initial state and final state configuration, whereas the HR factors quantify the electron-phonon coupling strength.

\begin{table}
\begin{center}
\caption{{H-X (X=Chalcogen atom) bond length for H-X configuration and H-M (M=Metal atom) bond length for H\textsubscript{$i$} configuration for all the studied TMDs. All the values are in {\AA}.}}
\begin{threeparttable}
\setlength{\tabcolsep}{6pt} 
\renewcommand{\arraystretch}{1.5} 
\begin{tabular}{lccc}
  \toprule\toprule
  \multirow{2}{*}{\raisebox{-\heavyrulewidth}{Material}} & \multicolumn{2}{c}{H-X bond length} & \multicolumn{1}{c}{H-M bond length}\\
  \cmidrule(lr){2-4}
   & neutral & +1 & +1 \\
  \midrule
  MoS\textsubscript{2} & 1.40 & 1.36 & 1.85 \\
  MoSe\textsubscript{2} & 1.57& 1.49 & 1.90 \\
  MoTe\textsubscript{2} & 1.84 & 1.70 & 1.98 \\
  WS\textsubscript{2} & 1.46 & 1.36 &  1.86\\
  WSe\textsubscript{2} & 1.59 & 1.49 & 1.92 \\
   \bottomrule\bottomrule
\end{tabular}
\label{table:bondlength}
\end{threeparttable}
\end{center}
\end{table}

Hydrogen is one of the most important impurities commonly found during the growth of TMDs. Zhen \textit{et al.} \cite{zhenhydrogen2016} showed that H$_2$ molecule does not form a covalent bond with MoS$_2$ and is electrically inactive with a migration barrier of 0.53 eV, which makes it highly unstable at high temperatures. In this work, we studied H$_i$ and H-X (X~=~chalcogen atom) as two possible configurations of H in TMDs. We find that for H$_i$, the hydrogen atom sits at the interstitial site in the plane of the metal atom inside the hexagonal lattice. In H-X configuration, the position of H depends upon the charge state. In the neutral charge state, the H-S bond is aligned with Mo-S bond pointing toward the empty space between the layers, whereas the H atom lies on top of the chalcogen for +1 charge state. Both configurations cause very minimal local relaxation. The H-X bond length in neutral and +1 charge states are 1.40~{\AA} and 1.36~{\AA}, respectively, in MoS$_2$, which is very close to the calculated bond length of 1.40~{\AA} in the H$_2$S molecule. The H-X bond length for all the other TMDs is reported in Table~\ref{table:bondlength}. For the H$_i$ configuration, the neighboring metal atoms which are in the same plane as H, are displaced outward away as a general trend for all studied TMDs. The H-M(M=metal atom) in-plane bond length for  MoS$_2$, MoSe$_2$, MoTe$_2$, WS$_2$ and WSe$_2$ for +1 charge state of the H$_i$ configuration is shown in Table~\ref{table:bondlength}. Singh \textit{et al.}~\cite{abhishek2019origin} discussed the interactions of hydrogen and native point defects (hydrogenated defects) in monolayer MoS$_2$ and found them to act as compensating centers in the presence of hydrogen.

In  Fig.~\ref{fig3}, we show the formation energy of the two H configurations for all the studied TMDs. Both configurations have the lowest formation energies in chalcogen-rich limit. H$_i$ configuration is more stable than H-X for the entire range of Fermi level for all the TMDs except for MoS$_2$, where H-X is slightly more stable in agreement with previous studies for MoS$_2$~\cite{singh2022}. H$_i$ acts as a shallow donor for all the TMDs and remains ionized in +1 charge state throughout the range of band gap except for MoS$_2$ and WS$_2$. The (+/0) transition levels are around 133 meV and 64 meV below the CBM for MoS$_2$ and WS$_2$, respectively. The H-X acts as a deep donor for all the studied TMDs with (+/0) transition level listed in Table~\ref{table:transition3}. In the metal-rich limit, the Fermi level will be pinned around 0.37 eV and 0.31 eV below the CBM for MoS$_2$ and WS$_2$, respectively. At these energies, the formation energy of H-S (H$_i$) in +1 charge and sulfur vacancy in -1 charge are equal for MoS$_2$ (WS$_2$). For all the other TMDs, H$_i$ acts as a shallow donor even in metal-rich limit and can lead to n-type conductivity.

\begin{table}
\begin{center}
\caption{(+/0) transition level for H-X configuration with respect to CBM for all the studied TMDs.}
\begin{threeparttable}
\setlength{\tabcolsep}{4pt} 
\renewcommand{\arraystretch}{1.5} 
\begin{tabular}{lc}
  \toprule\toprule
  \multirow{2}{*}{\raisebox{-\heavyrulewidth}{Material}} & \multicolumn{1}{c}{Transition level (eV) } \\
  \cmidrule(lr){2-2}
  & (+/0)  \\
  \midrule
  MoS\textsubscript{2} & 0.24    \\
  MoSe\textsubscript{2} & 0.34    \\
  MoTe\textsubscript{2} & 0.22  \\
  WS\textsubscript{2} & 0.08   \\
  WSe\textsubscript{2} & 0.42 \\
   \bottomrule\bottomrule
\end{tabular}
\label{table:transition3}
\end{threeparttable}
\end{center}
\end{table}

\begin{figure}
\includegraphics[width=8 cm]{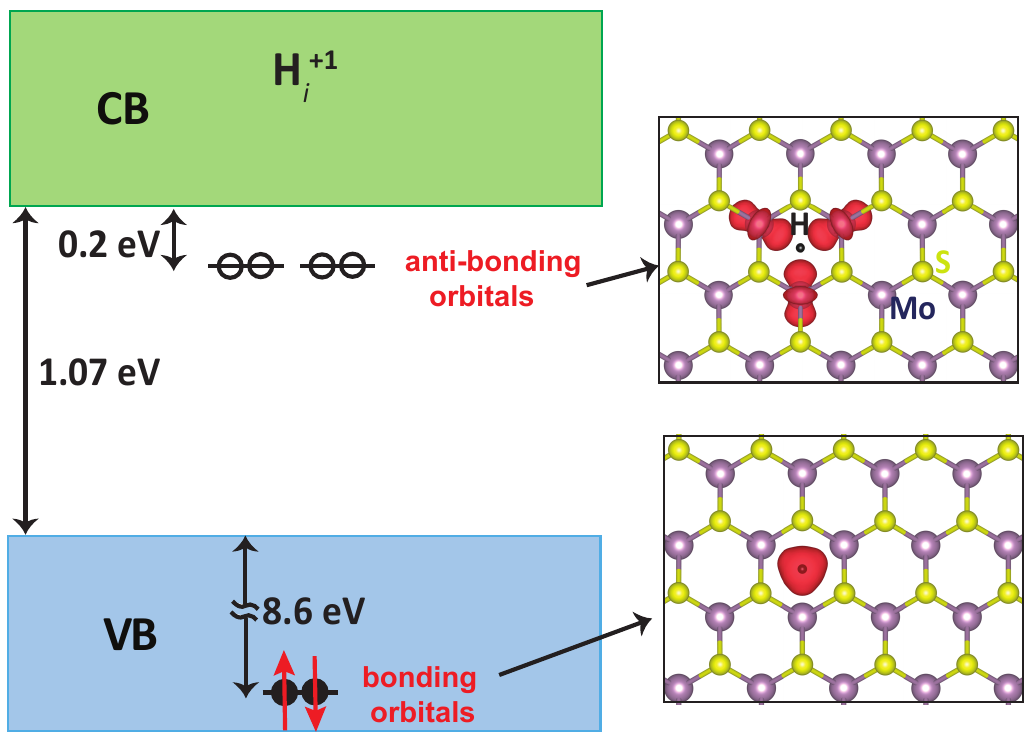}
\caption{Kohn-Sham defect states for H$_i$\textsuperscript{+1} configuration in bulk MoS$_2$. The fully occupied defect state is below the VBM and localized on the H atom, whereas the two empty states are in the band gap and localized on the neighboring Mo atom. The VB and CB correspond to the valence band and conduction band respectively.}
\label{fig5}
\end{figure}

\begin{table}
\begin{center}
\caption{{Vibrational frequencies of H$_i$\textsuperscript{+1} for all the TMDs. The in-plane modes are slightly different because of the different atomic environments around H$_i$\textsuperscript{+1}.}}
\begin{threeparttable}
\setlength{\tabcolsep}{4pt} 
\renewcommand{\arraystretch}{1.5} 
\begin{tabular}{lcccc}
  \toprule\toprule
  \multirow{2}{*}{\raisebox{-\heavyrulewidth}{Material}} & \multicolumn{2}{c}{Vibrational Frequencies ({cm\textsuperscript{-1}}) } \\
  \cmidrule(lr){2-3}
  & In-Plane modes & Out-of-Plane mode \\
  \midrule
  MoS\textsubscript{2} & 1550, 1526 & 603   \\
  MoSe\textsubscript{2} & 1331, 1311 & 665   \\
  MoTe\textsubscript{2} & 1028, 1021 & 691  \\
  WS\textsubscript{2} & 1580, 1545 & 744   \\
  WSe\textsubscript{2} & 1347, 1317 & 779 \\
   \bottomrule\bottomrule
\end{tabular}
\label{tab6}
\end{threeparttable}
\end{center}
\end{table}

In Fig.~\ref{fig5}, we show the electronic configuration of H$_i$ in MoS$_2$. When an H atom is placed at the center of the hexagon in the Mo plane, it interacts with neighboring three Mo atoms pushing them slightly outward. This interaction leads to the creation of three defect states. In the neutral charge state, one of the bonding-like fully occupied defect states lies around 8.28 eV below the VBM, whereas the other two anti-bonding defect states lie in the band gap around 0.13 eV below the CBM, one of these states is partially occupied with one electron. In the +1 charge state, the occupied bonding state lies around 8.6 eV below the VBM, and the two empty anti-bonding states are located around 0.2 eV below the CBM. The charge densities show that these defect states are highly localized. The bonding state has most of the contribution from the H $s$-orbital whereas the anti-bonding empty defect states are highly localized on neighboring Mo atoms. We also calculated the vibrational frequencies associated with H$_i$ in all the TMDs. There are three vibrational modes for the H$_i$ configuration; two of them are in-plane modes, and one mode is in the out-of-plane direction. The calculated frequencies listed in Table.~\ref{tab6} will help their identification. We also evaluated the stability of H$_i$ by calculating the migration barrier height of H$_i$\textsuperscript{+1} using nudged elastic band (NEB) method \cite{henkelman2000climbing}. The energy barriers of H$_i$\textsuperscript{+1} for migration either in the plane or out-of-plane lie in the range of 1.5 eV to 2.3 eV for all studied TMDs, making this defect quite stable at room temperature.

In summary, the chalcogen vacancy is a deep acceptor in all the studied bulk TMDs. The chalcogen vacancy can also lead to emission peaks near-infrared region as measured by a recent cathodoluminescence experiment for bulk MoS$_2$. The calculated emission peaks for all the studied bulk TMDs provide a guideline for probing chalcogen vacancy in future photoluminescence (PL)/cathodoluminescence (CL) experiments. We also investigated the possible configuration of H in bulk TMDs. Our results show that H$_i$ is a shallow donor in all the studied TMDs except MoS$_2$ and WS$_2$, and can lead to $n$-type conductivity in these materials. The Fermi level is pinned well below the CBM for MoS$_2$ and WS$_2$ in the metal-rich limit. The formation energy of H$_i$ is lower than H-X for most of the TMDs except MoS$_2$. 
We also determined the two in-plane and one of-plane frequency modes associated with H$_i$\textsuperscript{+1} for all the TMDs. Our results provide a detailed understanding of the chalcogen vacancy and the hydrogen impurity, providing a guideline for experiments to probe these defects in bulk TMDs.

This work was supported by the Laboratory Directed Research and Development (LDRD) Program (Grant No. PPPL-132)  at Princeton Plasma Physics Laboratory under U.S. Department of Energy Prime Contract No. DE-AC02-09CH11466. A.J. acknowledges funding from the National Science Foundation (NSF) award \#OIA-2217786. The calculations were carried out at the National Energy Research Scientific Computing Center (NERSC), a U.S. Department of Energy Office of Science User Facility located at Lawrence Berkeley National Laboratory, operated under Contract No. DE-AC02-05CH11231 using NERSC award BES-ERCAP27253, the Stellar Cluster at Princeton University, and the DARWIN computing system at the University of Delaware, using the NSF grant no.~1919839.

\bibliography{bulk_TMD}

\end{document}